\documentclass[twocolumn,prb,aps,nobalancelastpage,amsfonts]{revtex4}

\usepackage{graphicx}

\begin{document}

\title{Interpretation of the angular dependence of the de Haas-van Alphen effect in MgB$_2$}

\author{A. Carrington and J.D. Fletcher}

\affiliation{H.H. Wills Physics Laboratory, University of Bristol, Tyndall Avenue, BS8 1TL, United Kingdom.}

\author{H. Harima}
\affiliation{Department of Physics, Kobe University, Kobe, 657-8501 Japan.}

\date{\today}
\begin{abstract}
We present detailed results for the amplitude and field dependence of the de Haas-van Alphen (dHvA) signal arising from
the electron-like $\pi$ sheet of Fermi surface in MgB$_2$. Our data and analysis show that the dip in dHvA amplitude
when the field is close to the basal plane is caused by a beat between two very similar dHvA frequencies and not a
spin-zero effect as previously assumed.  Our results imply that the Stoner enhancement factors in MgB$_2$ are small on
both the $\sigma$ and $\pi$ sheets.
\end{abstract}

\maketitle

Soon after the discovery of superconductivity in MgB$_2$, several groups reported calculations of its band-structure.
The topology of the calculated Fermi surface was verified by measurement of the de Haas-van Alphen (dHvA) effect.  The
observed \cite{YellandCCHMLYT02,CarringtonMCBHYLYTKK03} dHvA frequencies were in very good agreement with the
calculated \cite{Harima02,MazinK02,RosnerAPD02} extremal areas on each of the four sheets of Fermi surface.  The
measured quasiparticle effective masses were also in excellent agreement with the calculated values, providing direct
evidence for the predicted large difference in electron-phonon coupling constants between the $\sigma$ and $\pi$ bands.

For fields less than 20~T, three dHvA frequencies have much larger amplitude than the others; two of these arise from
the minimal and maximal extremal areas of the smaller of the two tubular $\sigma$ sheets, whereas the third originates
from a neck on the electron-like $\pi$ sheet.  In what follows we refer to these dHvA orbits as $F_1$, $F_2$ and $F_3$
respectively (a diagram showing the calculated Fermi surface and predicted dHvA orbits can be found in Ref.\
\onlinecite{CarringtonMCBHYLYTKK03}).

\begin{figure}
\includegraphics[width=8.0cm,keepaspectratio=true]{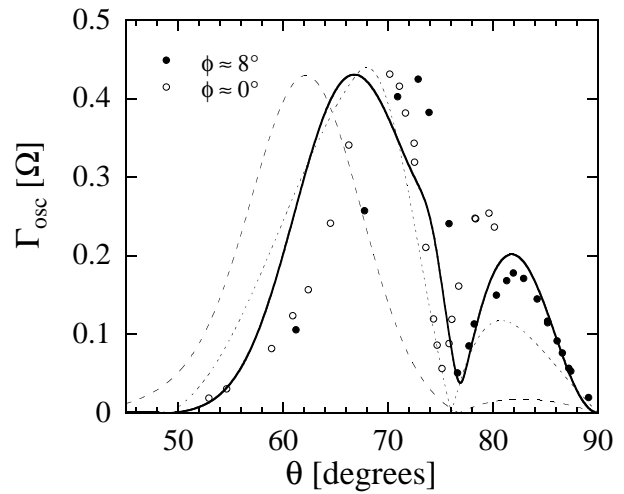}
 \caption{Experimental dHvA torque amplitude versus $\theta$ at fixed field ($17.8~T<B<18.0~T$) for orbit $F_3$. Data for two
 different in-plane rotation angles $\phi$ are shown. Fits to the two frequency model [Eq.\ (\ref{eqbeat})], warped
 cylinder model [Eq.\ (\ref{eqry})] and single frequency model [Eq.\ (\ref{eqlk})]
  are shown by the solid, dotted and dashed  lines respectively.}
 \label{figamptheta}
\end{figure}

In Ref.\ \onlinecite{CarringtonMCBHYLYTKK03}, it was shown that the angular dependence of the dHvA amplitude for $F_1$
and $F_2$ could be adequately explained by the usual Lifshitz-Kosevich expression for the oscillatory torque
$\Gamma_{osc}$ of a 3D Fermi liquid,\cite{shoenberg} but for $F_3$ it could not. There were two puzzling features.
First, the amplitude of $F_3$ showed a pronounced dip at $\theta\simeq 76^\circ$ (see Fig.\
\ref{figamptheta})\cite{angledef} which was attributed to a `spin-zero' effect. The Stoner enhancement deduced from the
position of this dip is approximately two times larger than that predicted\cite{MazinK02} and four times larger than
those measured for the $\sigma$ sheet orbits, $F_1$ and $F_2$. Second, we were unable to explain the angle dependence
of the amplitude. In this paper we will present detailed measurements of the field dependent dHvA amplitude of $F_3$ as
a function of angle, and show that the feature previously ascribed to a `spin-zero' is actually caused by a beat with
another dHvA frequency.

Quantum oscillations in the torque produced by a small single crystal of MgB$_2$, (mass=5.6$\mu$g) were measured with a
piezoresistive, doped silicon cantilever.\cite{piezolevers}  Changes in the resistance of the cantilever are directly
proportional to the torque and were measured using an AC bridge technique.\cite{CooperCMYHBTLKK03} The torque values
are reported here in units of bridge resistance $R$, i.e, the off-balance voltage divided by the excitation current
(the change in lever resistance is 4 times larger than this).  We estimate that $\Gamma=10^{-10}$R ($\Gamma$ in Nm and
$R$ in $\Omega$). The noise level is around 2m$\Omega$ or $\sim 10^{-13}$ Nm. The crystal of MgB$_2$ was grown by a
high pressure synthesis route \cite{angledef} and is the same as sample B in Refs.\ \onlinecite{YellandCCHMLYT02} and
\onlinecite{CarringtonMCBHYLYTKK03}.

We interpret our data using the standard expression for the first harmonic of the oscillatory part of the torque for a
3D Fermi liquid \cite{shoenberg,mistake}
\begin{equation}
\Gamma_{osc}\propto
\frac{B^\frac{3}{2}}{[\mathbb{A}^{\prime\prime}]^\frac{1}{2}}\frac{dF}{d\theta}R_DR_TR_{SC}R_S\sin\left[\frac{2\pi
F}{B}+\varphi\right]
\label{eqlk}
\end{equation}
where $F$ is the dHvA frequency [$F = (\hbar/2\pi e)\mathbb{A}$,  $\mathbb{A}$ is the extremal orbit area in
$\mathbf{k}$-space]; $\mathbb{A}^{\prime\prime}=\partial^2\mathbb{A}/\partial k^2$ is the curvature factor and
$\varphi$ is the phase. $R_D$, $R_T$ and $R_S$ are the damping factors from impurity scattering, temperature and spin
splitting respectively. The Dingle factor, $R_D=\exp(-\frac{\pi \hbar k_F}{ eB\ell})$, where $k_F$ is the orbitally
averaged Fermi wavevector \cite{shoenberg,WassermanS94} and $\ell$ is the quasiparticle mean free path. The thermal
damping factor, $R_T = X/\sinh X$ where $X = (2\pi^2k_{_B}m^*T)/(\hbar eB)$, $m^*$ is the quasi-particle effective
mass. The factor $R_{SC}$ accounts for the additional damping when the sample enters the superconducting state, and was
studied in detail in MgB$_2$ in Ref.\ \onlinecite{FletcherCKK03}. The spin splitting factor is given by $R_S = \cos[(
\pi g m_B[1+S])/(2m_e)]$ where $1+S$ is the orbitally averaged exchange-correlation (Stoner) enhancement factor, $g$ is
the electron $g$-factor, $m_{e}$ is the free-electron mass. When $g (1+S)(m_B/m_e)$ equals an integer, $R_s=0$ and the
spin-up and spin-down Fermi surfaces beat out of phase to produce a `spin-zero' minimum in the dHvA amplitude. If the
location of any of these `spin-zeros' can be measured, and $m_B(\theta,\phi)$ is known from band-structure
calculations, then the Stoner factor on each orbit may be deduced.  Note that $R_s$ does not depend on $B$, so near to
a `spin-zero' the dHvA amplitude is suppressed at all fields.

Although it is possible to fit the data directly to Eq.\ (\ref{eqlk}) it is often more illuminating to extract the
field dependent amplitude $A$ of the dHvA oscillations by fitting small sections of the data (comprising of 1.5
oscillations) to
\begin{equation}
\Gamma_{osc}(B)=A\sin\left(\frac{2\pi F}{B}+\varphi\right)+aB+b
\end{equation}
(the linear term accounts for the slowly varying background torque and magnetoresistance of the cantilever). We then
divide $A$ by $B^\frac{3}{2} R_T$ to give the reduced amplitude $\tilde{A}$, which in the absence of other effects
(i.e., $R_{SC}=1$)\cite{Rscnote} is proportional to the Dingle factor $R_D$. The quasiparticle effective mass $m^*$ in
the expression for $R_T$ was determined by measuring the temperature dependence of the dHvA amplitude [for $F_3$, $m^*
= (0.456 \pm 0.005) m_e$ at $\theta = 70.8^\circ$].

\begin{figure}
\includegraphics[width=8.0cm,keepaspectratio=true]{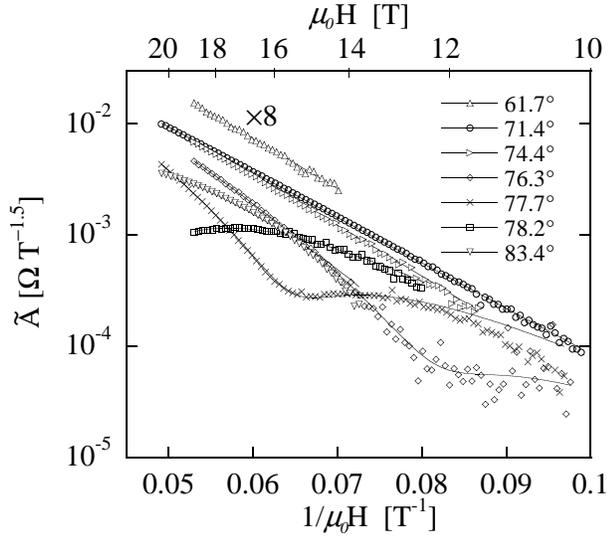}
 \caption{Reduced dHvA amplitude ($\tilde{A}$) versus inverse field for several rotation angles. The lines are fits of
 the data to Eq.\ (\protect\ref{eqbeat}). The data for $\theta=61.7^\circ$ have been multiplied by 8 for clarity.}
 \label{figdingle}
\end{figure}

In Fig.\ \ref{figdingle} we show reduced dHvA amplitude $\tilde{A}$ versus inverse field (`Dingle' plot) for orbit
$F_3$ at selected angles.  For $\theta=61.7^\circ$ and $71.4^\circ$ the behavior is strictly exponential,
$\tilde{A}\propto \exp(-\alpha/B)$. However,  for $\theta \gtrsim 71.4^\circ$, $\alpha$ \textit{appears} to increase
rapidly and the Dingle curves become markedly non-exponential. The dHvA amplitude is strongly reduced, but unlike the
expected behavior close to a `spin-zero', the reduction is not uniform at all fields. For $\theta<72^\circ$ the
increase in the coefficient $\alpha$ with decreasing angle is given by $\alpha= (89/\sin\theta) \times
\epsilon(\theta)$. The $\sin\theta$ factor arises from the cylindrical nature of this section of Fermi surface (the
band mass and the dHvA frequency increase by the same factor). The factor $\epsilon(\theta)$ was determined
experimentally by fitting the data for $\theta<72^\circ$; we find, $\epsilon(\theta)\simeq 1+3.7(1-\sin\theta)^2$
reflecting an increase in scattering rate with decreasing $\theta$. From this we estimate the quasiparticle mean free
path (at $\theta=0^\circ$) on this orbit to be 660~\AA.

We will show below that the most likely reason for the non-exponential Dingle curves is a beat with a second dHvA
frequency.  Before we present this analysis in detail we will briefly discuss two other possible explanations.

The background and oscillatory torque cause deflections in the cantilever so the measurements are not performed at
strictly constant angle.  This `torque interaction' effect \cite{BergemannJMTFMN99} causes spurious generation of
harmonics and mixing of frequencies, and because the amplitude is angle dependent it can also cause bending of the
Dingle curves.  We can quantify this effect by fitting the data to  Eq.\ (\ref{eqlk}), allowing for the change in dHvA
frequency with field. To a good approximation\cite{YellandCCHMLYT02} $F_3(\theta)=F_3^0/\sin(\theta)$, and $\theta$ is
given by the measured angle plus the field dependent deflection of the cantilever, $\theta=\theta_0+\lambda\Gamma(B)$
($\lambda$ is a fitting constant which measures the `stiffness' of the cantilever). For the data presented here we find
that $\lambda\simeq0.04$ degrees/$\Omega$.\cite{leverstiffness}  The result of this analysis is that the torque
interaction effect is much too small to explain the non-exponential behavior see in Fig.\ \ref{figdingle}.

Another possibility stems from a mosaic spread in the direction of the $a$ or $c$ axis of the crystal. Beats between
different dHvA frequencies originating from different part of the crystal can produce non-exponential Dingle
curves.\cite{higgins} The size of the effect depends on the spread of the dHvA frequencies and in the present case
would becomes largest as the field is rotated away from basal plane. Experimentally we find that curvature in the
Dingle plot only occurs for angles \textit{close} to the basal plane, and therefore conclude that in our crystals the
mosaic structure does not play a significant role.  This conclusion is reinforced by the fact that the observed
behavior is highly reproducible between different crystals.

\begin{figure}
\includegraphics[width=8.0cm,keepaspectratio=true]{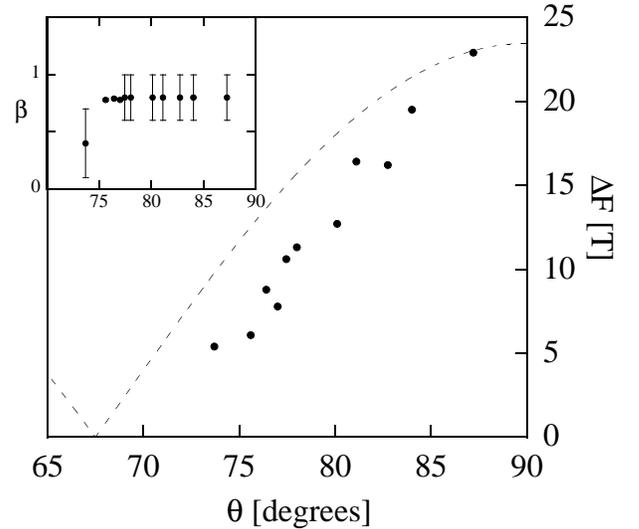}
 \caption{Variation of the frequency difference $\Delta F$ and the relative amplitude $\beta$ of the beat frequency
 as a function of $\theta$.  The dashed line is Eq.\ (\ref{eqyam}), $l^*=0.12\frac{a^*}{\sqrt{3}}$ and $\Delta F_0=23.5~$T.}
 \label{figdfamp}
\end{figure}

Simple trigonometric relations show that if two dHvA signals  with frequencies $F$ and $F+\Delta F$ and amplitudes in
the ratio $\beta$ (with identical Dingle factors) are added together, and the resultant fitted with a single frequency
$F$ then the reduced amplitude $\tilde{A}$ will vary like
\begin{eqnarray}
\tilde{A}&=&\tilde{A_0}\exp\left(-\frac{\alpha}{B}\right)\mathbb{D}(B)\sin\left(\frac{2\pi F}{B}+\delta(B)\right)\nonumber\\
\mathbb{D}(B)&=&\left[1+\beta^2+2\beta\cos\left(\frac{2\pi\Delta F}{B}\right)\right]^\frac{1}{2}\nonumber\\
\delta(B)&=&\tan^{-1}\left[\frac{\beta \sin\left(\frac{2\pi\Delta F}{B}\right)}{1+\beta\cos\left(\frac{2\pi\Delta
F}{B}\right)}\right]. \label{eqbeat}
\end{eqnarray}
In Fig.\ \ref{figdingle} we show that these equations provide a very good fit to the reduced dHvA amplitude data.  In
these fits there are three free parameters, $\tilde{A_0}$, $\Delta F$ and $\beta$ ($\alpha$ was fixed at the values
found by extrapolating from lower $\theta$ as described above).  The variations of $\Delta F$ and $\beta$ with $\theta$
are shown in Fig.\ \ref{figdfamp}. The frequency difference increases approximately linearly with $\theta$, $\Delta F
\simeq 1.4 (\theta-70^\circ)$. The data are consistent with $\beta$ being almost constant as a function of angle and
then going rapidly to zero for $\theta\lesssim 73^\circ$.  Note that as $F_3(\theta)\simeq2684/\sin(\theta)$, the
maximum value of $\Delta F$ corresponds to only  $\sim 0.4\%$ difference in $k_F$ between the two orbits.

\begin{figure}
  \includegraphics[width=6cm]{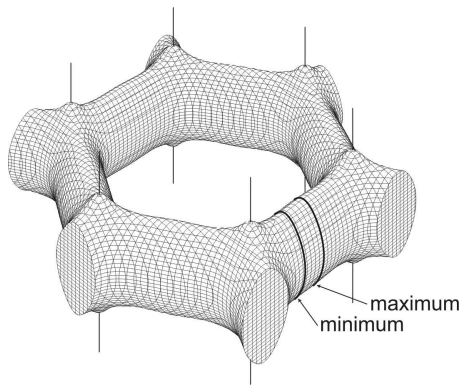}
  \caption{The electron-like $\pi$ band Fermi surface sheet of MgB$_2$, showing the location of the two `$F_3$' dHvA orbits.}\label{f3fig}
\end{figure}

A extremal dHvA orbit $F_3^*$ with frequency close to $F_3$ was predicted from band structure calculations
\cite{Harima02} and is shown in Fig. \ref{f3fig}.  Close to the two extremal orbits this sheet of Fermi surface is
tubular, with only slight warping. The difference in dHvA frequency between the extremal orbits of a cylinder with
simple cosine warping\cite{Yamaji89,YoshidaMSMIOBHMAS99} is given by
\begin{equation}\label{dfeq}
  \Delta F(\theta)=\frac{\Delta F_0}{\sin(\theta)} J_0\left[\frac{\pi k_F}{l^*} \cot(\theta)\right]
  \label{eqyam}
\end{equation}
Here $l^*$ is the  $k$-space distance between the minimum and maximum frequency orbits and $J_0$ is the Bessel
function.  For $\theta\gtrsim 65^\circ$, the detailed band structure determination of $\Delta F_3 = F_3^*-F_3$ closely
follows this equation with $\Delta F_0\simeq57~$T, and $l^*\simeq 0.13 \frac{a^*}{\sqrt{3}}$ (Ref.\
\onlinecite{notea}). For $\theta\lesssim 65^\circ$, $\Delta F_3(\theta)=0$, i.e., only one extremal orbit is found.
This reflects a difference between the actual Fermi surface topology and the cosine dispersion which leads to Eq.\
(\ref{eqyam}). The characteristics of this predicted second orbit closely match our observations. The only significant
difference is that the maximum frequency difference we find is around half the predicted value (i.e., the cylinder is
actually somewhat less warped than the calculation). Experimentally, the Dingle curves are straight for $\theta
\lesssim 72^\circ$ (Fig.\ \ref{figdingle}), showing that the second orbit is absent for $\theta\lesssim 72^\circ$ in
approximate agreement with the detailed calculation.

We conclude that the dip in amplitude for orbit $F_3$ is caused by a beat effect and not a spin-zero as previously
assumed.  From the data in Fig.\ \ref{figamptheta} it seems likely that any spin zero would have to occur for
$\theta<50^\circ$ and hence [assuming  for $F_3$, $m_B\propto(\sin\theta)^{-1}$] the Stoner enhancement factor on this
orbit is less than 0.22.   A summary of all the Stoner factors derived from dHvA measurements is shown in Table
\ref{stable} (the values for $F_1$ and $F_2$ are taken from a previous study\cite{CarringtonMCBHYLYTKK03} on a
different MgB$_2$ crystal\cite{whyxtlk}).  The enhancement factors on both the $\sigma$ and $\pi$ sheets are somewhat
smaller than calculations\cite{MazinK02} would suggest. By comparing measurements of the Pauli susceptibility, derived
from conduction electron spin resonance experiments, to band-structure calculations of the total density of states at
the Fermi level, Simon \textit{et al.}\cite{SimonJFMGFPBLKC01} found that $1+S$ (averaged over all Fermi surface
sheets)was $0.86\pm0.13$, i.e., the average enhancement is small in agreement with our findings.

\begin{table}
\begin{ruledtabular}
\caption{Summary of measured ($1+S_{\rm exp}$) and calculated \protect\cite{MazinK02} ($1+S_{\rm calc}$) Stoner
enhancement factors for the three orbits.} \label{stable}
\begin{tabular}{llll}
Orbit&Sheet&$1+S_{\rm exp}$&$1+S_{\rm calc}$\\
$F_1$ & $\sigma_1$ & 1.07 & 1.31 \\
$F_2$ & $\sigma_1$ & 1.12 & 1.31 \\
$F_3$ & $\pi_1$ & $<1.22$ & 1.26 \\
\end{tabular}
\end{ruledtabular}
\end{table}

Using the measured values of $\alpha$, $\Delta F(\theta)$, and  $\beta(\theta)$  along with the calculated curvature
factor $\mathbb{A}^{\prime\prime}$, we are able to calculate the expected angular dependence of the dHvA amplitude at
fixed field using Eq.\ (\ref{eqlk}).  Here we have fixed the constant of proportionality in Eq.\ (\ref{eqlk}) to match
the maximum in the experimental data, and we have also fixed $1+S=1.2$. As mentioned above, $1+S$ must be $\lesssim
1.22$ and values less than $1.2$ do not change the curve markedly. The agreement between the calculation and the data
is good, although not perfect.  Importantly, the calculation reproduces the key dip feature at $\theta\simeq76^\circ$.

Actually, for such a flat portion of Fermi surface, the usual LK expression [Eq.\ (\ref{eqlk})] should be modified as
the usual integral over the tube is no longer strongly peaked at the extremal orbits.  For a simple cosine dispersion
the curvature factor is replaced by\cite{YoshidaMSMIOBHMAS99}
\begin{equation}\label{eqry}
  R_Y=J_0\left[\frac{\pi \Delta F(\theta)}{B}\right].
\end{equation}
A fit to the amplitude data with Eqs. (\ref{dfeq}) and (\ref{eqry}) is shown as the dotted line in Fig.\
\ref{figamptheta}.  Here we have set $l^*=0.12\frac{a^*}{\sqrt{3}}$ and $\Delta F_0=23.5~T$ to best fit both the
$\Delta F(\theta)$ and $\Gamma_{osc}(\theta)$ data. The fit again reproduces the main features of the experimental data
but is slightly worse than the discrete frequency fit.  It can be seen (Fig.\ \ref{figdfamp}) that the experimentally
determined $\Delta F_3(\theta)$ data does not quite follow Eq.\ (\ref{eqyam}), showing that the true local warping is
not quite cosinusodial. The behavior below $\theta=74^\circ$, where the second frequency is not detected
experimentally, is particularly badly described by this approximation.

The dashed line in Fig.\ \ref{figamptheta} shows the behavior expected without the beat effect but with $1+S=1.545$,
which is required to produce a `spin-zero' dip at the same angle. Clearly the correspondence with the data is much
worse, especially for angles close to the basal plane.

For the series of sweeps reported in Figs.\ \ref{figdingle} and \ref{figdfamp}, the in-plane angle $\phi$ was estimated
to be $\sim8^\circ$.  A second (less extensive) set of runs with $\phi\simeq 0^\circ$ shows very similar behavior. The
main difference is that the dip occurs at lower $\theta$ (see Fig.\ \ref{figamptheta}). Repeating the above analysis on
this set of data again shows a linear $\theta$ dependence of $\Delta F$ but with a larger slope, $d\Delta
F/d\theta\simeq$ 1.9 T/degree. If we use this value in our calculation we find the dip moves to lower $\theta$, in
accord with the experimental data.

In summary, we have shown that the dip in the dHvA amplitude for fields aligned close to the basal plane of MgB$_2$ is
due to a beat between two very similar dHvA frequencies and not a spin-zero effect as previously assumed.  The data
imply that the Stoner enhancement factors on both the $\sigma$ and $\pi$ sheets of Fermi surface are small.

We thank J.R.\ Cooper for useful discussions and S.\ Lee and S.\ Tajima for supplying the single crystals of MgB$_2$.

\bibliographystyle{apsrevNOETAL}
\bibliography{mgb2F3zero}

\begin{thebibliography}{21}
\expandafter\ifx\csname natexlab\endcsname\relax\def\natexlab#1{#1}\fi
\expandafter\ifx\csname bibnamefont\endcsname\relax
  \def\bibnamefont#1{#1}\fi
\expandafter\ifx\csname bibfnamefont\endcsname\relax
  \def\bibfnamefont#1{#1}\fi
\expandafter\ifx\csname citenamefont\endcsname\relax
  \def\citenamefont#1{#1}\fi
\expandafter\ifx\csname url\endcsname\relax
  \def\url#1{\texttt{#1}}\fi
\expandafter\ifx\csname urlprefix\endcsname\relax\def\urlprefix{URL }\fi
\providecommand{\bibinfo}[2]{#2}
\providecommand{\eprint}[2][]{\url{#2}}

\bibitem[{\citenamefont{Yelland et~al.}(2002)\citenamefont{Yelland, Cooper,
  Carrington, Hussey, Meeson, Lee, Yamamoto, and Tajima}}]{YellandCCHMLYT02}
\bibinfo{author}{\bibfnamefont{E.~A.} \bibnamefont{Yelland}},
  \bibinfo{author}{\bibfnamefont{J.~R.} \bibnamefont{Cooper}},
  \bibinfo{author}{\bibfnamefont{A.}~\bibnamefont{Carrington}},
  \bibinfo{author}{\bibfnamefont{N.~E.} \bibnamefont{Hussey}},
  \bibinfo{author}{\bibfnamefont{P.~J.} \bibnamefont{Meeson}},
  \bibinfo{author}{\bibfnamefont{S.}~\bibnamefont{Lee}},
  \bibinfo{author}{\bibfnamefont{A.}~\bibnamefont{Yamamoto}}, \bibnamefont{and}
  \bibinfo{author}{\bibfnamefont{S.}~\bibnamefont{Tajima}},
  \bibinfo{journal}{Phys. Rev. Lett.} \textbf{\bibinfo{volume}{88}},
  \bibinfo{pages}{217002} (\bibinfo{year}{2002}).

\bibitem[{\citenamefont{Carrington et~al.}(2003)\citenamefont{Carrington,
  Meeson, Cooper, Balicas, Hussey, Yelland, Lee, Yamamoto, Tajima, Kazakov, and
  Karpinski}}]{CarringtonMCBHYLYTKK03}
\bibinfo{author}{\bibfnamefont{A.}~\bibnamefont{Carrington}},
  \bibinfo{author}{\bibfnamefont{P.~J.} \bibnamefont{Meeson}},
  \bibinfo{author}{\bibfnamefont{J.~R.} \bibnamefont{Cooper}},
  \bibinfo{author}{\bibfnamefont{L.}~\bibnamefont{Balicas}},
  \bibinfo{author}{\bibfnamefont{N.~E.} \bibnamefont{Hussey}},
  \bibinfo{author}{\bibfnamefont{E.~A.} \bibnamefont{Yelland}},
  \bibinfo{author}{\bibfnamefont{S.}~\bibnamefont{Lee}},
  \bibinfo{author}{\bibfnamefont{A.}~\bibnamefont{Yamamoto}},
  \bibinfo{author}{\bibfnamefont{S.}~\bibnamefont{Tajima}},
  \bibinfo{author}{\bibfnamefont{S.~M.} \bibnamefont{Kazakov}},
  \bibnamefont{and}
  \bibinfo{author}{\bibfnamefont{J.}~\bibnamefont{Karpinski}},
  \bibinfo{journal}{Phys. Rev. Lett.} \textbf{\bibinfo{volume}{91}},
  \bibinfo{pages}{037003} (\bibinfo{year}{2003}).

\bibitem[{\citenamefont{Harima}(2002)}]{Harima02}
\bibinfo{author}{\bibfnamefont{H.}~\bibnamefont{Harima}},
  \bibinfo{journal}{Physica C} \textbf{\bibinfo{volume}{378}},
  \bibinfo{pages}{18} (\bibinfo{year}{2002}).

\bibitem[{\citenamefont{Mazin and Kortus}(2002)}]{MazinK02}
\bibinfo{author}{\bibfnamefont{I.~I.} \bibnamefont{Mazin}} \bibnamefont{and}
  \bibinfo{author}{\bibfnamefont{J.}~\bibnamefont{Kortus}},
  \bibinfo{journal}{Phys. Rev. B} \textbf{\bibinfo{volume}{65}},
  \bibinfo{pages}{180510} (\bibinfo{year}{2002}).

\bibitem[{\citenamefont{Rosner et~al.}(2002)\citenamefont{Rosner, An, Pickett,
  and Drechsler}}]{RosnerAPD02}
\bibinfo{author}{\bibfnamefont{H.}~\bibnamefont{Rosner}},
  \bibinfo{author}{\bibfnamefont{J.~M.} \bibnamefont{An}},
  \bibinfo{author}{\bibfnamefont{W.~E.} \bibnamefont{Pickett}},
  \bibnamefont{and} \bibinfo{author}{\bibfnamefont{S.~L.}
  \bibnamefont{Drechsler}}, \bibinfo{journal}{Phys. Rev. B}
  \textbf{\bibinfo{volume}{66}}, \bibinfo{pages}{024521}
  (\bibinfo{year}{2002}).

\bibitem[{\citenamefont{Shoenberg}(1984)}]{shoenberg}
\bibinfo{author}{\bibfnamefont{D.}~\bibnamefont{Shoenberg}},
  \emph{\bibinfo{title}{Magnetic Oscillations in Metals}}
  (\bibinfo{publisher}{Cambridge University Press},
  \bibinfo{address}{Cambridge}, \bibinfo{year}{1984}).

\bibitem[{ang()}]{angledef}
\bibinfo{note}{Here, $\theta$ is the angle between the field and the $c$-axis
  as the sample is rotated towards the basal plane. The in-plane direction we
  are rotating towards is denoted by $\phi$ (measured from the $a$-axis).}

\bibitem[{pie()}]{piezolevers}
\bibinfo{note}{Veeco Instruments Inc., Woodbury, New York;
  \url{http://www.veeco.com}}.

\bibitem[{\citenamefont{Cooper et~al.}(2003)\citenamefont{Cooper, Carrington,
  Meeson, Yelland, Hussey, Balicas, Tajima, Lee, Kazakov, and
  Karpinski}}]{CooperCMYHBTLKK03}
\bibinfo{author}{\bibfnamefont{J.~R.} \bibnamefont{Cooper}},
  \bibinfo{author}{\bibfnamefont{A.}~\bibnamefont{Carrington}},
  \bibinfo{author}{\bibfnamefont{P.~J.} \bibnamefont{Meeson}},
  \bibinfo{author}{\bibfnamefont{E.~A.} \bibnamefont{Yelland}},
  \bibinfo{author}{\bibfnamefont{N.~E.} \bibnamefont{Hussey}},
  \bibinfo{author}{\bibfnamefont{L.}~\bibnamefont{Balicas}},
  \bibinfo{author}{\bibfnamefont{S.}~\bibnamefont{Tajima}},
  \bibinfo{author}{\bibfnamefont{S.}~\bibnamefont{Lee}},
  \bibinfo{author}{\bibfnamefont{S.~M.} \bibnamefont{Kazakov}},
  \bibnamefont{and}
  \bibinfo{author}{\bibfnamefont{J.}~\bibnamefont{Karpinski}},
  \bibinfo{journal}{Physica C} \textbf{\bibinfo{volume}{385}},
  \bibinfo{pages}{75} (\bibinfo{year}{2003}).

\bibitem[{mis()}]{mistake}
\bibinfo{note}{Note that in Ref.\ \onlinecite{CarringtonMCBHYLYTKK03} the
  exponent of $B$ in Eq.\ \ref{eqlk} is wrongly given as $\frac{1}{2}$.}

\bibitem[{\citenamefont{Wasserman and Springford}(1994)}]{WassermanS94}
\bibinfo{author}{\bibfnamefont{A.}~\bibnamefont{Wasserman}} \bibnamefont{and}
  \bibinfo{author}{\bibfnamefont{M.}~\bibnamefont{Springford}},
  \bibinfo{journal}{Physica B} \textbf{\bibinfo{volume}{194}},
  \bibinfo{pages}{1801} (\bibinfo{year}{1994}).

\bibitem[{\citenamefont{Fletcher et~al.}(2004)\citenamefont{Fletcher,
  Carrington, Kazakov, and Karpinski}}]{FletcherCKK03}
\bibinfo{author}{\bibfnamefont{J.~D.} \bibnamefont{Fletcher}},
  \bibinfo{author}{\bibfnamefont{A.}~\bibnamefont{Carrington}},
  \bibinfo{author}{\bibfnamefont{S.~M.} \bibnamefont{Kazakov}},
  \bibnamefont{and}
  \bibinfo{author}{\bibfnamefont{J.}~\bibnamefont{Karpinski}},
  \bibinfo{journal}{Phys. Rev. B} \textbf{\bibinfo{volume}{70}},
  \bibinfo{pages}{144501} (\bibinfo{year}{2004}).

\bibitem[{Rsc()}]{Rscnote}
\bibinfo{note}{As discussed in Ref.\ \onlinecite{FletcherCKK03}, the data in
  Fig.\ \ref{figdingle} are taken at sufficiently high field so that the effect
  of superconductivity can be neglected (expect possibly for
  $\theta=77.7^\circ$ and $1/\mu_0H\gtrsim 0.09$).}

\bibitem[{\citenamefont{Bergemann et~al.}(1999)\citenamefont{Bergemann, Julian,
  Mackenzie, Tyler, Farrell, Maeno, and Nishizaki}}]{BergemannJMTFMN99}
\bibinfo{author}{\bibfnamefont{C.}~\bibnamefont{Bergemann}},
  \bibinfo{author}{\bibfnamefont{S.~R.} \bibnamefont{Julian}},
  \bibinfo{author}{\bibfnamefont{A.~P.} \bibnamefont{Mackenzie}},
  \bibinfo{author}{\bibfnamefont{A.~W.} \bibnamefont{Tyler}},
  \bibinfo{author}{\bibfnamefont{D.~E.} \bibnamefont{Farrell}},
  \bibinfo{author}{\bibfnamefont{Y.}~\bibnamefont{Maeno}}, \bibnamefont{and}
  \bibinfo{author}{\bibfnamefont{S.}~\bibnamefont{Nishizaki}},
  \bibinfo{journal}{Physica C} \textbf{\bibinfo{volume}{318}},
  \bibinfo{pages}{444} (\bibinfo{year}{1999}).

\bibitem[{lev()}]{leverstiffness}
\bibinfo{note}{We find that this parameter can vary by up to a factor 4 between
  different cantilevers from the same supplier.}

\bibitem[{\citenamefont{Higgins and Lowndes}(1980)}]{higgins}
\bibinfo{author}{\bibfnamefont{R.~J.} \bibnamefont{Higgins}} \bibnamefont{and}
  \bibinfo{author}{\bibfnamefont{D.~H.} \bibnamefont{Lowndes}}, in
  \emph{\bibinfo{booktitle}{Electrons at the Fermi surface}}, edited by
  \bibinfo{editor}{\bibfnamefont{M.}~\bibnamefont{Springford}}
  (\bibinfo{publisher}{Cambridge University Press}, \bibinfo{year}{1980}).

\bibitem[{\citenamefont{Yamaji}(1989)}]{Yamaji89}
\bibinfo{author}{\bibfnamefont{K.}~\bibnamefont{Yamaji}}, \bibinfo{journal}{J.
  Phys. Soc. Jpn.} \textbf{\bibinfo{volume}{58}}, \bibinfo{pages}{1520}
  (\bibinfo{year}{1989}).

\bibitem[{\citenamefont{Yoshida et~al.}(1999)\citenamefont{Yoshida, Mukai,
  Settai, Miyake, Inada, Onuki, Betsuyaku, Harima, Matsuda, Aoki, and
  Sato}}]{YoshidaMSMIOBHMAS99}
\bibinfo{author}{\bibfnamefont{Y.}~\bibnamefont{Yoshida}},
  \bibinfo{author}{\bibfnamefont{A.}~\bibnamefont{Mukai}},
  \bibinfo{author}{\bibfnamefont{R.}~\bibnamefont{Settai}},
  \bibinfo{author}{\bibfnamefont{K.}~\bibnamefont{Miyake}},
  \bibinfo{author}{\bibfnamefont{Y.}~\bibnamefont{Inada}},
  \bibinfo{author}{\bibfnamefont{Y.}~\bibnamefont{Onuki}},
  \bibinfo{author}{\bibfnamefont{K.}~\bibnamefont{Betsuyaku}},
  \bibinfo{author}{\bibfnamefont{H.}~\bibnamefont{Harima}},
  \bibinfo{author}{\bibfnamefont{T.~D.} \bibnamefont{Matsuda}},
  \bibinfo{author}{\bibfnamefont{Y.}~\bibnamefont{Aoki}}, \bibnamefont{and}
  \bibinfo{author}{\bibfnamefont{H.}~\bibnamefont{Sato}}, \bibinfo{journal}{J.
  Phys. Soc. Jpn.} \textbf{\bibinfo{volume}{68}}, \bibinfo{pages}{3041}
  (\bibinfo{year}{1999}).

\bibitem[{not()}]{notea}
\bibinfo{note}{$a^*$ is the in-plane reciprocal lattice vector and
  $\frac{a^*}{\sqrt{3}}$ is the length of one side of the hexagonal Brillouin
  zone.}

\bibitem[{why()}]{whyxtlk}
\bibinfo{note}{For the present crystal the scattering rate on the $\sigma$
  sheet orbits is too large to observe the dHvA signal at sufficiently large
  angle for the `spin-zero' to be observed.}

\bibitem[{\citenamefont{Simon et~al.}(2001)\citenamefont{Simon, Janossy, Feher,
  Muranyi, Garaj, Forro, Petrovic, Bud'ko, Lapertot, Kogan, and
  Canfield}}]{SimonJFMGFPBLKC01}
\bibinfo{author}{\bibfnamefont{F.}~\bibnamefont{Simon}},
  \bibinfo{author}{\bibfnamefont{A.}~\bibnamefont{Janossy}},
  \bibinfo{author}{\bibfnamefont{T.}~\bibnamefont{Feher}},
  \bibinfo{author}{\bibfnamefont{F.}~\bibnamefont{Muranyi}},
  \bibinfo{author}{\bibfnamefont{S.}~\bibnamefont{Garaj}},
  \bibinfo{author}{\bibfnamefont{L.}~\bibnamefont{Forro}},
  \bibinfo{author}{\bibfnamefont{C.}~\bibnamefont{Petrovic}},
  \bibinfo{author}{\bibfnamefont{S.~L.} \bibnamefont{Bud'ko}},
  \bibinfo{author}{\bibfnamefont{G.}~\bibnamefont{Lapertot}},
  \bibinfo{author}{\bibfnamefont{V.~G.} \bibnamefont{Kogan}}, \bibnamefont{and}
  \bibinfo{author}{\bibfnamefont{P.~C.} \bibnamefont{Canfield}},
  \bibinfo{journal}{Phys. Rev. Lett.} \textbf{\bibinfo{volume}{8704}},
  \bibinfo{pages}{047002} (\bibinfo{year}{2001}).

\end{thebibliography}

\end{document}